\documentstyle[12pt]{article}
\def\p{\partial}
\def\bq{\begin{eqnarray}}
\def\eq{\end{eqnarray}}

\begin{document}
\title{\bf Non-renormalizability of a SSB gauge theory
induced by a non-linear fermion-Higgs coupling of canonical dimension 4}
\author{NISTOR NICOLAEVICI\\
\it Technical University of Timi\c soara, Department of Physics,\\
\it P-\c ta Hora\c tiu 1, RO-1900 Timi\c soara, Romania}
\maketitle
\begin{abstract}
We consider an abelian gauge theory with spontaneously broken symmetry 
containing a scalar-fermion coupling which is non-linear in the Higgs field. 
Although 
in the unitary gauge it reduces to a pure Yukawa term, suggesting that the 
theory is renormalizable, the one loop divergence structure in this gauge 
in the fermion-fermion elastic scattering amplitude shows this is not 
so. Comparison is made with the theory with a conventional coupling, for which 
cancellation of the non-renormalizable divergences occurs.
\end{abstract}

Some years ago it was proposed in Ref. \cite{cotaescu} a class of pure left 
SU(N)$\otimes$U(1) gauge theories with spontaneously broken symmetry, capable
of accommodating the Weinberg-Salam model and some of its 
generalizations\footnote{See discussions in Ref. 1.} up to the Higgs sector. 
They exhibit the unfamiliar feature of using a scalar-fermion coupling 
$non$-$linear$ in the Higgs fields, specially designed, however, to reduce 
to a Yukawa term in the unitary gauge. It is natural to ask whether 
such a construction is generally compatible with 
renormalizability. We show here, by pointing to a simple abelian model, that 
renormalizability can be lost.

Consider the U(1) theory given by the Lagrangian 
${\cal L}={\cal L}_1+{\cal L}_2$, with the 
first term corresponding to the Goldstone model $(\mu^2, \lambda>0)$
\bq
{\cal L}_1&=&-\frac{1}{4}(\p_\mu A_\nu-\p_\nu A_\mu)^2+
(D_\mu \Phi)^+(D^\mu \Phi)
+\frac{\mu^2}{2}\Phi^+\Phi-\lambda(\Phi^+\Phi)^2,\label{l1}
\eq
and 
\bq
{\cal L}_2&=&i\bar \psi \gamma^\mu D_\mu\psi-G\,\bar\psi \psi \sigma,\quad
D_\mu=\p_\mu-ig A_\mu,\quad G>0,
\eq
where the $\sigma$ field is defined by writing
\bq
\Phi=\sigma e^{i\chi},
\eq
with $\sigma$, $\chi$ real. The fermion-Higgs coupling is of the type 
mentioned above: it is not linear in $\Phi$, while in the unitary gauge 
$\chi=0$ it reduces to a genuine Yukawa term.

The non-linearity associated with $\Phi$ makes unitary gauge a most 
convenient choice for writing the Feynman rules. Let $m_f$, 
$M$ denote the masses (in the tree approximation) acquired by the fermion and 
vectorial field after symmetry breaking, and let 
$\frac{\varphi}{\sqrt{2}}$ represent the shifted $\sigma$ field with zero vacuum 
expectation value. Then the interaction Lagrangian is 
\bq
{\cal L}_{int}=
-g\bar \psi\gamma^\mu\psi A_\mu+gM \varphi A^\mu A_\mu +\frac{1}{2}g^2
\varphi^2 A^\mu A_\mu\\ - g\frac{m_f}{M}\bar \psi\psi \varphi+
g\frac{\mu^2}{2M}\varphi^3+g^2\frac{\mu^2}{M^2}\varphi^4,
\label{lint}
\eq
and the Feynman rules can be read off from the effective interaction 
Hamiltonian \cite{weinberg}
\bq 
{\cal H}_{eff}=-{\cal L}_{int}+i\delta^4(0)\ln
\left(1+g\frac{\varphi}{M}\right),
\label{heff}
\eq
using the covariant vector field propagator 
$D_{\mu\nu}(k)=(-g_{\mu \nu}+k_\mu k_\nu/M^2)/(k^2-M^2+i\epsilon)$.

Now, it is 
generally accepted \cite{unitary} that despite the bad ultraviolet 
behaviour of the massive vectorial propagators in unitary gauge, 
renormalizability should still allow the $S$ matrix divergences to be absorbed 
into a proper set of counterterms (which is not the case for arbitrary 
Green functions). When calculating physical quantities, the hidden gauge 
symmetry is expected to lead in to the cancellation of the non-renormalizable 
divergences in individual graphs. It turns out this fails to happen for our 
theory.

Consider the $g^4$ contribution to the fermion-fermion elastic
scattering. The corresponding Feynman graphs are displayed in the
Figure. We have not shown (i) diagrams which differ only by an external line 
permutation, (ii) tadpole and external line self-energy diagrams 
(as taken care by mass and wave function renormalization), and (iii) 
diagrams involving only $\psi-\varphi$ interactions. The last ones can be 
easily seen to contain only renormalizable divergences.

The result of a straightforward calculation can be summarized as follows. The 
following pairs of diagrams are finite, though each diagram is separately 
divergent: (a, b), (c, d), (e, f), (g, h), (i, j).
Diagrams (k), (l) prove to be individually finite. Vertex diagrams (m), (n) are
logarithmically, while (o), (p), (q) and (r) quadratically divergent. The
divergences turn out to be momentum-independent, so that in principle they 
can be cured
by coupling constant (eventually $m_f$, $M$) renormalization. Quadratical 
divergences in diagrams (s), (t) correspond to vector field mass and 
wave function renormalization. One is left with the quartically 
divergent self-energy graph (u). Its contribution reads
\bq
f=2g^4m_f^2
\bar u(p_f)u(p_i)\bar u(q_f) u(q_i)\frac{\Sigma(k)}{(k^2-\mu^2)^2},
\eq
where $k=p_f-p_i=q_i-q_f$ and
\bq
\Sigma(k)=\int \frac{d^4q}{(2\pi)^4}
\frac{g_{\mu \nu}-q_\mu q_\nu/M^2}
{q^2-M^2+i\epsilon}\,\,
\frac{g^{\mu \nu}-(k^\mu-q^\mu)(k^\nu-q^\nu)/M^2}
{(k-q)^2-M^2+i\epsilon}.
\label{sig}
\eq
Evaluating the integral in $n=4-\epsilon$ dimensions \cite{tHooft} one gets 
for the divergent part
\bq
\Sigma_\infty (k)=\frac{i}{32 \pi^2\epsilon}
\frac{(k^2)^2}{M^4}+\frac{3i}{8\pi^2\epsilon},
\eq
which leads, along with mass and wave function renormalization terms, to 
the non-renormalizable four-fermion interaction 
\bq
f^{NR}_\infty=\frac{ig^4}{16\pi^2\epsilon}
\left(\frac{m_f}{M^2}\right)^2
\bar u(p_f)u(p_i)\bar u(q_f) u(q_i).
\eq
This suffices to prove our point.

It is interesting to consider the same process in the theory modified to
allow a conventional coupling. It is not hard to see that the only way to do 
this without changing the particle spectrum (or destroying relativistic 
invariance) is to resort to the familiar choice with chiral 
components of $\psi$ behaving differently under gauge transformations. 
One can set, without loss of generality,
\bq
\psi_L\rightarrow e^{-ig\Lambda}\psi_L, \quad \psi_R\rightarrow \psi_R,
\eq
where $\psi_L$, $\psi_R$ are the left and right handed projections, 
respectively. This does not affect ${\cal L}_1$, but maintaining U(1) gauge 
invariance requests replacing ${\cal L}_2$ with
\bq
{\cal L}_2^\prime=i\bar \psi \gamma^\mu (\p_\mu-ig\frac{1-\gamma_5}{2}A_\mu)
\psi-G\,((\bar\psi_L \Phi)\psi_R+\bar\psi_R (\Phi^+\psi_L)).
\label{l2prime}
\eq

The resulting ${\cal L}^\prime={\cal L}_1+{\cal L}_2^\prime$ theory is just 
the Appelquist and Quinn model \cite{appel} used in the early days of 
electroweak theory to test divergence cancellations in unitary gauge
calculations. In this gauge, the $S$ matrix still follows from 
the effective Hamiltonian (\ref{heff}) with the sole modification 
in ${\cal L}_{int}$ concerning the spinor-vector coupling
\bq
\bar \psi \gamma^\mu \psi A_\mu\,
\rightarrow\,
\bar \psi\gamma^\mu\frac{1-\gamma_5}{2}\psi A_\mu.
\label{aint}
\eq
This proves essential for divergence cancellations. 
Situation now presents as follows (the contributing Feynamn graphs in both 
theories are obviously the same). Diagrams (a) and 
(b) summed yield a logarithmic divergence which equals $f_\infty^{NR}$. 
Diagrams (k) and (l) produce logarithmic and quadratically divergences which 
added yield, for each diagram, $-f^{NR}_{\infty}$. These four diagrams thus 
exactly cancel the divergence in diagram (u). The rest of diagrams 
prove to contain either renormalizable infinities, or non-renormalizable ones 
corresponding to a pseudoscalar-pseudoscalar 
four-fermion interaction, which again cancel among 
themselves\footnote{See Ref. 5. Cancellation of
the one-loop non-renormalizable divergences was also checked in Ref. 6, though 
not by a direct evaluation of individual graphs.}

In conclusion, our example shows one should be cautious in extending the 
renormalizability result \cite{tHooft1} concerning 
SSB gauge theories when including in the fermion-scalar terms non-linearities 
in the Higgs sector, even though by formal power counting this might look 
justified. 

\bigskip
\noindent
\section*{Acknowledgments}
I am indebted to Ion I. Cot\u aescu and Geza Ghemes for stimulating discussions and
criticisms. I also thank Atilla Farkas for reading the manuscript.

\end{document}